# Terahertz semiconductor laser source at -12 $^0$C

Ali Khalatpour[1†*], Man Chun Tam[2*], Sadhvikas J. Addamane[3], John Reno[3], Zbig Wasilewski[2], Qing Hu[1†]

[1]*Department of Electrical Engineering and Research Laboratory of Electronics, Massachusetts Institute of Technology, Cambridge, MA 02139, USA;* [2]*Department of Electrical and Computer Engineering, University of Waterloo, 200 University Ave W, Waterloo, Ontario N2L 3G1, Canada.*

[3]*Sandia National* Laboratories, *Center of Integrated Nanotechnologies, Albuquerque, NM 87185-130, USA.*

*Equal contributions

[†]Corresponding authors: akhalatpour@g.harvard.edu, qhu@mit.edu

**Abstract**

Room temperature operation of Terahertz Quantum Cascade Lasers (THz QCLs) has been a long-pursued goal to realize compact semiconductor THz sources. The progress toward high-temperature operation in THz QCLs has been relatively slow compared to infrared QCLs owing to more significant challenges at THz frequencies. Recently, the maximum operating temperature of THz QCLs was improved to 250 K, and the achievement revitalized hope in the THz community in pursuit of higher temperature operations. In this paper, we report on further improvement in operating temperature to ~261 K (-12 $^0$C) by judiciously optimizing key parameters and discuss the challenges ahead in achieving room temperature operation.

# 1. Introduction

The THz spectral range (0.5 -10 THz) lies in the gap between frequency ranges accessible with conventional semiconductor electronic and photonic devices. Scientific curiosity and the abundance of THz applications in imaging [1] and spectroscopy [2] have motivated researchers to take up the challenge and explore various sources to fill this so-called "THz gap". At the lower end of the THz gap, electronic devices such as transistors and frequency multipliers work well, but their power levels drop precipitously above ~1 THz. With carefully engineered power combining circuits, frequency multipliers have reached ~1.6 THz with milliwatt (mW) optical power at room temperature [3,4]. Both quantum cascade laser-pumped molecular lasers (QPMLs) [5] and difference-frequency generation (DFG) [6,7] have used IR QCLs to generate THz waves at room temperature indirectly. Quantum cascade lasers (QCLs) have several advantages compared to the nonlinear frequency upconversion and downconversion schemes. As fundamental oscillators, THz QCLs inherently have much higher power levels and efficiencies than nonlinear frequency conversions. At Cryogenic temperatures, THz QCL systems can reach watt-level optical power in pulse mode [8-10] and tens of mW in continuous wave operation with ~ 1% efficiency [11,12]. In addition, THz QCL gain medium can be used to develop novel THz sources such as frequency combs [13,14] and radiation amplifiers [15]. Another advantage of THz QCLs is enabling novel cavities with unique radiation properties. For instance, the local oscillator at 4.74 THz for GUSTO [16], which is a NASA Balloon Born mission, was developed using a tunable unidirectional cavity that achieves ~10 mW of CW power with less than 1.8 W dissipated power, and it was cooled with a compact Stirling cooler (CryoTel®CT) as light as 5 Kg [17]. The required size for cooling and the system's vibration will be significantly reduced with thermoelectric (TE) coolers.

Recently, compact hand-held TE-cooled THz QCLs have been demonstrated in which an average power of ~150 $\mu$ W enabled real-time imaging with a THz camera[18]. This achievement was made possible by a record high operating temperature of $T_{max}$ ~ 250 K. In principle, even a higher averaged power can be collected by increasing the duty cycle, using a high reflective coating of the back facet of the laser ridge, and collimating the output beam by silicon lens from the front facet. Such measures can potentially increase the collected output power by ~10 folds [19]. We project that TE-cooled THz QCLs with tens of mW average output power can be realized with a further increase of the maximum operating temperature. Here, we report a new record $T_{\max} = 261 \text{ K}$.

## 2. Toward a higher operating temperature in THz QCLs

Fig. 1(a) shows a schematic of a three-level laser to illustrate the dominant scattering processes in THz QCLs [20]. Here the upper lasing level, the lower lasing level, and the ground states in the $n^{th}$ module are denoted by $|u_n\rangle$, $|l_n\rangle$, $|g_n\rangle$ respectively. *IFR*, *imp*, *e-e*, *LO*$_{em}$, and *LO*$_{abs}$ stand for interface roughness scattering, impurity scattering, electron-electron scattering, *LO*-phonon emission, and *LO*-phonon absorption, respectively. In this scheme, electrons are injected into $|u_n\rangle$ and extracted from $|l_n\rangle$ through resonant tunneling and scattering with *LO*-phonons in the heterostructure. The peak gain of intersubband transitions $G_p$ can be described by the oscillator strength $f_{ul}$, the population inversion between the upper and the lower lasing level $\Delta N$, and the transition linewidth $\Delta \nu$ as $G_p \propto \dfrac{\Delta N f_{ul}}{\Delta \nu}$. The linewidth is defined as $\Delta \nu = \dfrac{1}{2\pi}\left(\dfrac{1}{\tau^u} + \dfrac{1}{\tau^l} + \dfrac{2}{T^*}\right)$ in which $\tau^u$, $\tau^l$, $T^*$ are the upper-state lifetime, the lower-state lifetime, and pure dephasing times. The oscillator strength is essentially a wavefunction overlap integral between the upper and lower lasing levels and it is defined as $f_{ul} = (\dfrac{2m^*\Delta_E}{\hbar^2})|\langle u|z|l\rangle|^2$ in which $z$ is the growth direction, $m^*$ is the effective mass, and $\Delta_E$ is the energy separation between the two levels. The explicit temperature dependence of $\Delta N$ arises from thermally activated, non-radiative *LO*-phonon emission [20] and the so-called thermal backfilling due to *LO*-phonon absorption from $|g_n\rangle$ to $|l_n\rangle$ [21]. Although the form factors for the emission of photon and *LO*-phonon are different, they both strongly depend on the overlap of wavefunctions between $|u_n\rangle$ and $|l_n\rangle$ [20]. Hence, suppression of *LO*-phonon emission from the upper state comes at the cost of reducing the optical gain. Consequently, designs with higher $f_{ul}$ (vertical designs) have shorter $\tau^u$ than those with a lower $f_{ul}$ (diagonal designs). In diagonal designs, higher carrier concentration (higher doping) can be used to compensate for the reduced optical gain. However, increasing the doping leads to an increase in $\Delta \nu$ due to the reduction of $\tau^u$ and $T^*$ due to nonradiative elastic processes, including *e-e* and *imp* scattering. Therefore, the figure of the merit $\frac{f_{ul}\tau^u}{\Delta \nu}$ appears to limit the performance of diagonal designs targeting high temperatures. [22].

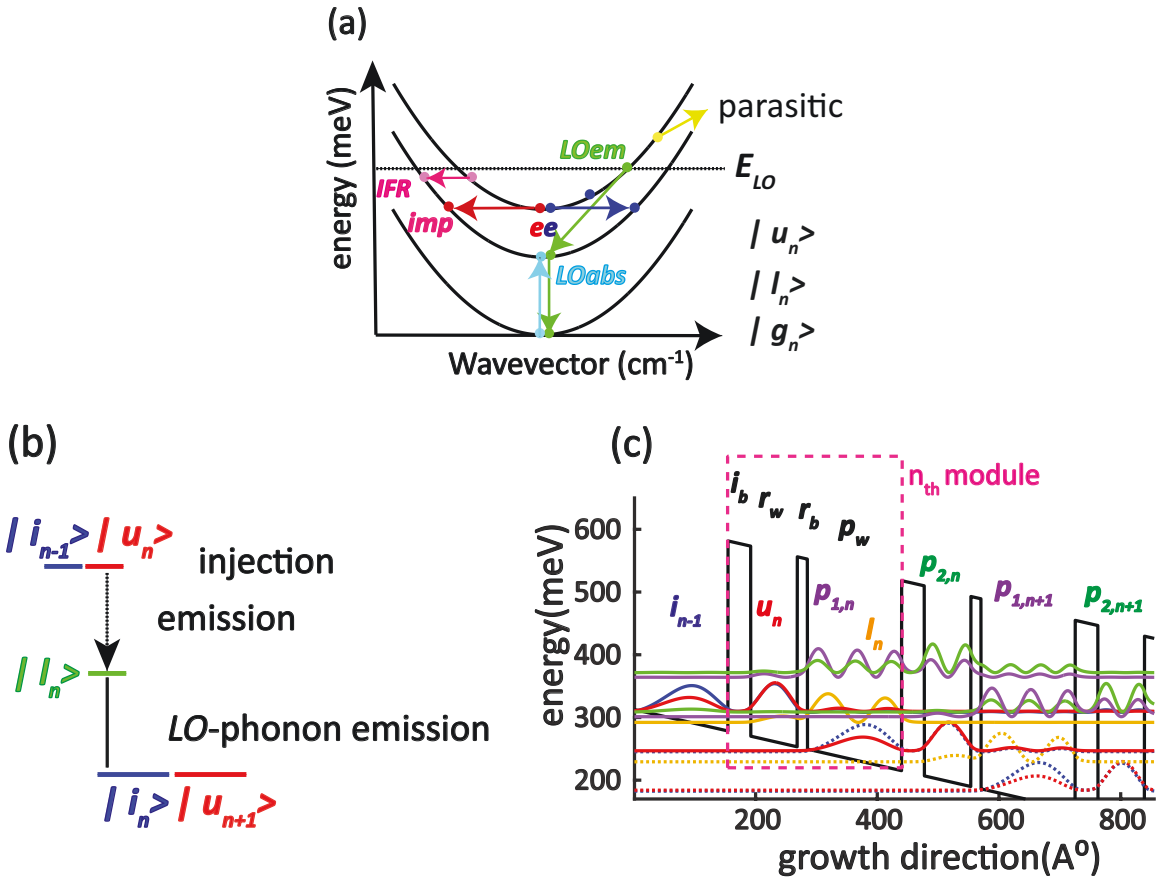

Fig.1. (a) schematic of a three-level QCL system. (b) a simple schematic of energy levels in a THz QCLs based on the direct phonon depopulations. (c) probability density functions of subband states in a THz QCL with a two-wells scheme.

Meanwhile, the escape of electrons to the continuum bands was postulated as an unaccounted mechanism in THz QCLs for their temperature degradation [23]. To counteract the escape to the continuum, higher barrier compositions were used in THz QCL designs [18, 23, 24]. To reduce the increased impact of *IFR* scattering—which scales quadratically with barrier composition—the number of barriers involved in electron transport should be minimized to reduce $\Delta v$. The most straightforward design in this direction is the so-called two-well scheme based on direct phonon depopulation, as illustrated in Fig.1(b) [25,26].

Fig. 1(c) shows the probability density functions of subband states in a THz QCL with a two wells scheme. Here The injection barrier, radiation well, radiation barrier, and phonon well are labeled, $i_b, r_w, r_b, p_w$ respectively. Due to a shorter module length in this scheme, undesired coupling between neighboring modules may occur at the injection alignment, $|i_{n-1}\rangle \to |u_n\rangle$. The coupling between $|i_{n-1}\rangle$, $|u_n\rangle$ and the bound states in the neighboring modules, $|p_{1,n}\rangle$, $|p_{2,n}\rangle$, $|p_{1,n+1}\rangle$, $|p_{2,n+1}\rangle$ was investigated, and by suppressing them, a record-high $T_{max} = 250$ K was achieved [18]. The energy separation between $|i_{n-1}\rangle$ and $|p_{1,n}\rangle$ (denoted as $E_{i,pp}$) can be made arbitrarily high to reduce a possible leakage from $|i_{n-1}\rangle$ to $|p_{1,n+1}\rangle$ [18]. However, as discussed in the supplemental file, the suppression of intermodule leakage (by increasing the $E_{i,pp}$) places an unfavorable upper limit on $f_{ul}\tau^u_{LO-em}$, where $\tau^u_{LO-em}$ is the lifetime due to nonradiative *LO*-phonon emission. In addition, if intermodule leakage is not suppressed, the transport simulations using a non-equilibrium green function (NEGF) solver predicted results that substantially deviated from experiments.

**3. Optimization, experimental results, and discussions for the two-well scheme**

In THz QCLs, a helpful graph to show the coupling between various channels is an anticrossing graph. An example of such a graph for several designs considered in this section is shown in Fig. 2(a). This graph shows the anticrossing curves over a wide range of module biases. Minima in the anticrossing correspond to a resonant alignment of two levels in energy, and the anticrossing gap is a measure of the coupling strength between those two levels. Ideally, the injection anticrossing (the minima in solid line) that corresponds to alignment, $|i_{n-1}\rangle \rightarrow |u_n\rangle$ should be close to the dephasing rate. The dephasing rate can be measured (through the radiation linewidth of spontaneous emission) and is ~4 meV in THz QCL heterostructures. The transparency of the injection barrier ensures the minimum required doping to achieve population inversion and reduces the linewidth broadening due to *imp* scattering [22]. However, increasing the transparency of the injection barrier comes at the cost of an increase in the parasitic channels corresponding to the coalignment between $|i_{n-1}\rangle$, $|u_n\rangle$ and the bound states in the neighboring modules, $|p_{1,n}\rangle, |p_{2,n}\rangle, |p_{1,n+1}\rangle, |p_{2,n+1}\rangle$ ( minima in the dot lines)[18]. Here we optimized $\zeta = \dfrac{\Omega_{i_{n-1},u_n}}{\Omega_{u_n,p_{1,n+1}}}$ as a figure of merit to achieve both objectives. In semiconductor lasers, such as THz QCLs, the rate of increase in the threshold current density ($J_{th}$) with temperature carries information on the leakage into the parasitic channels [23,27]. In addition, $J_{max} - T$ can be used to evaluate the transparency of the injection barrier. In a so-called lifetime limited transport, where the maximum current density through the injection barrier is limited by the upper state lifetime, $J_{max}$ is temperature dependent. On the other hand, in tunneling-limited transport, the $J_{max}$ is mainly limited by dephasing processes and is temperature-independent [28,29]. A slight decrease in $J_{max}$ with temperature in diagonal designs indicates lifetime limited transport [30] and is sought after in this manuscript. The simulated gain vs. temperature using NEGF is shown in Fig. 2(b) for these designs. The maximum operating temperature $T_{max}$ can be inferred from Fig. 2(b), at which the gain decreases to the total cavity loss of ~20 cm$^{-1}$ [31]. This was corroborated by the temperature performance of G652 with $T_{max} \approx 250$ K [18]. Further details on the optimization scheme can be found in the supplemental file. A summary of growth parameters, the maximum operating temperature, and maximum and threshold current densities are summarized in Table 1.

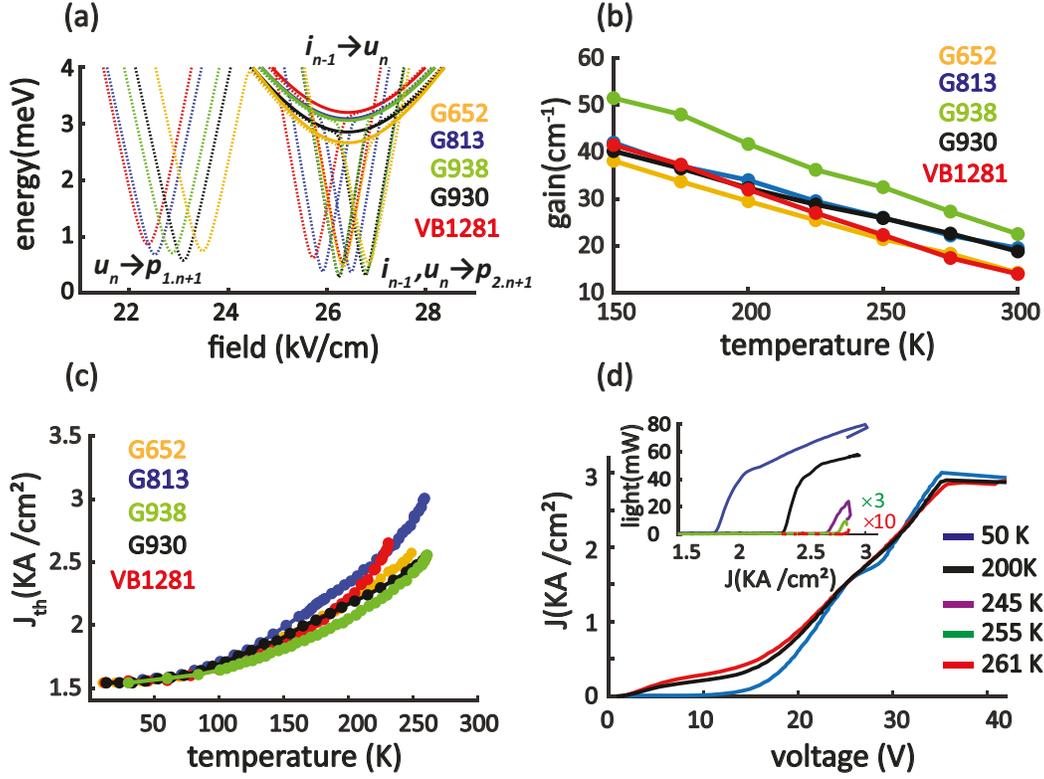

Fig.2. (a) anticrossing plot for various suggested designs. (b) simulated gain vs. temperature using NEGF (c) experimentally measured $J_{th} - T$. The values of $J_{th}$ from multiple devices are scaled to that of G652 at 10 K for easy comparison. (d) The voltage-current density- light characterization of G938. The sudden slope change in $L$-$J$ relations at T ≤ 200 K is due to detector saturation. The lasing frequency is ~ 4 THz.

A 25% Al barrier was used in VB1281 to reduce the impact of *IFR* scattering. As are shown in Fig 2(b). the NEGF simulations predict a similar gain in VB1281 compared to G652 with a $T_{max}$ =250K [18]. The $T_{max}$ =232 K was achieved for this structure, the highest reported for designs based on 25% barriers (previously 210 K [24]). Since no drop of $J_{max} - T$ was observed in VB1281 (suggesting a tunneling transport regime), some improvements in $T_{max}$ may still be possible using a higher $\zeta$. However, as shown in Fig 2(c), a higher rate of increase in $J_{th} - T$ in VB1281 was measured as compared to G652. This higher rate suggests thermal leakage to parasitic channels is more severe in 25% Al barriers and could explain the overestimation of gain in NEGF simulations.

Table 1. The layer sequence starts from the injection barrier. Bold denotes barriers separating GaAs quantum wells. The phonon well is indicated by a parenthesis, and the red ink indicates the doped region. The volume doping n$_{3D}$ is defined relative to that of G652 (1.5e$^{17}$ cm$^{-3}$) and represented as $n_{3D}^{G652}$ in the table. Here, x represents the aluminum fraction in the barriers. The number of modules is chosen for ~10 $\mu$m of an active region.

| label | x (±0.003) | $T_{max}$ (±0.1 k) | $J_{th}^{20K} - J_{max}^{T_{max}}$ (± 20 A/cm$^2$) | Layer sequence (Å) (±.5 Å) | Doping |
|---|---|---|---|---|---|
| VB1281 | 0.25 | 232 | 1600-2650 | **38.4**, 69.7, **24.2**, (93,25,27) | $n_{3D}^{G652}$ |
| G813 | 0.3 | 260 | 1350-2600 | **32.7**, 70.4, **20.8**, (57,30,57) | $n_{3D}^{G652}$ |
| G930 | 0.35 | 259 | 1450-2700 | **30.4**, 72.5, **19.1**, (59,28,59) | $1.2\, n_{3D}^{G652}$ |
| G902 | 0.35 | 260 | 2050-3350 | **30.6** 73.0, **19.0**,(56,35,55) | $1.2\, n_{3D}^{G652}$ |
| G938 | 0.35 | 261 | 1750-2850 | **28.8**, 74.5, **17.6**,(30,30,90) | $n_{3D}^{G652}$ |

In designs based on 30% Al barriers, the $T_{max}$ for G813 was improved to 260 K. However, a higher rate of increase in $J_{th} - T$ compared to G652 was measured, as shown in Fig.2(c). Judging from the anticrossing plot in Fig 2(a) and the simulated subband population for G813 using NEGF shown in Fig. S7 (d) of the supplemental file, no apparent leakage channels are introduced compared to G652. Here, the regrowth of G813 showed the same behavior so growth problems can be ruled out. One possible explanation for the higher rate of increase in $J_{th} - T$ could be the $|l_n\rangle \to |p_{2,n+1}\rangle$ alignment. Further explorations are required to shed more light on the impact of the $|l_n\rangle \to |p_{2,n+1}\rangle$ alignment on the behavior of $J_{th} - T$. Though not shown here, the $J_{max} - T$ of G813 showed a negligible decrease with temperature, suggesting that the transport is tunneling limited and further reduction of injection barrier thickness and lower $f_{ul}$ (a higher $\zeta$) is worth pursuing. The details of this design labeled as VA1186-B are presented in the supplementary file.

Another direction to improve the $T_{max}$ of G813 is through doping scheme optimization, as explained in detail in the supplementary file.

In designs with 35% barriers, a similar temperature performance was measured in G930 as compared to G813, which is consistent with the NEGF simulations shown in Fig. 2(b). However, as shown in Fig. 2(c), the rate of increase of $J_{th}$ with temperature is reduced in G930 compared to G813 and G652. This implies a higher suppression of thermal leakages. However, since G930 has a higher barrier composition, the *IFR* scattering rate might have been understated in our model for this barrier composition. Motivated by a simulated improvement of ~10 K in $T_{max}$, a design labeled as G902 with an identical design as G930 but with higher doping (X1.2) was explored. However, no further improvement was observed even though G902 benefited from a higher growth quality than G930. This could be explained by an underestimation of broadening due to *imp* and *e-e* scattering in the simulation. In G902, the measured $J_{max}$ decreased from 3400 A/cm² to 3350 A/cm² from 10 K to 260 K suggesting the onset of a lifetime limited transport. Therefore, further increase of $\zeta$ in this design was not pursued.

As discussed in the supplemental file, designs with a wider phonon-well have a higher value of the figure of merit $f_{ul}\tau_{LO-em}^{u}$ and show a higher gain in the simulations—however, designs based on this resulted in an unfavorable fast increase in $J_{th}-T$. In G938, we explored the possibility of a wider phonon-well than G930. This design also has a similar $\zeta$ as G813, which makes the comparison with the 30% designs easier. The doping position was also optimized in G938 using the model explained in the supplementary information file. As shown in Fig. 2(b), the simulated gain in G938 is higher than G930 when the same statistical parameters are used to describe the interface. However, the $T_{max}$ was not improved as much as expected. The rate of $J_{th}-T$ in G938 did not increase as compared to G930, in contrast to the trend observed in 30% barriers for wider phonon wells. Two possible experimental factors might have negatively affected the $T_{max}$ in G938. First, the fabrication of G938 experienced multiple problems, and only the wafer edge pieces were used for its characterization. We often see ~5-10 K improvement using centerpieces compared to edge pieces in our experiments. Second, the MBE growth quality of G938 is poorer than G930 as shown in the supplementary file information. Future re-growth of this wafer could verify if the improvement predicted by simulation for G938 compared to G930 can be experimentally achieved. The measured Voltage-Current- Light for G938 is shown in Fig. 2(d). The maximum current density in G938 is reduced from 3000 A/cm² at 10 K to 2850 A/cm² at 261 K. Therefore, further increase of $\zeta$ in this design was not pursued. However, co-aligning $|l_n\rangle \to |p_{2,n+1}\rangle$ and $|i_{n-1}\rangle \to |u_n\rangle$ is worth pursuing in 35% of barriers to explore if a double depopulation scheme can result in a higher gain.

**4. Exploring direct phonon depopulation with two injector wells**

Through the investigation so far, intermodule leakage places various restrictions on the two-well schemes. One possible approach to relax these limits is to use direct phonon depopulation with multiple injector wells. Fig. 3(a) shows an example of such design with two injectors, $^1i_{n-1}$ and $^2i_{n-1}$. For simplicity, we refer to this scheme as a three-well scheme. From the simulation point of view, with the same barrier composition, an optimized three-well scheme has a lower gain than an optimized two-well design. This can be qualitatively explained as a three-well scheme requires higher doping than a two-well scheme to reach the same $J_{max}$ because of a longer module length. As explained earlier, higher doping results in more gain broadening. In addition, as more barriers are involved in the injection to the upper lasing level, this scheme's dephasing by *IFR* scattering is also higher. However, there are still two important potential advantages to pursuing the three-well schemes. In perspective, transport simulations predict that room temperature operation could be achieved by reducing the injection barrier thickness and increasing doping (increased $J_{max}$) in all our designs (such as VB1281 and G813). However, these predictions were not experimentally realized. In other words, if the simulation assumptions are more valid in a three-well scheme because of reduced intermodule leakages, then its simulated $T_{max}$ will be closer to the experimental results than that for the two-well designs, even though the latter may have a higher predicted $T_{max}$. Optimization based on a more reliable model is always preferred. Another benefit of using a three-well scheme, which requires lower fields in the bias, is the possibility of using lower barrier composition to reduce *IFR* scattering. As shown in Fig. S6(c) in the supplemental file, *IFR* scattering significantly reduces achievable gains.

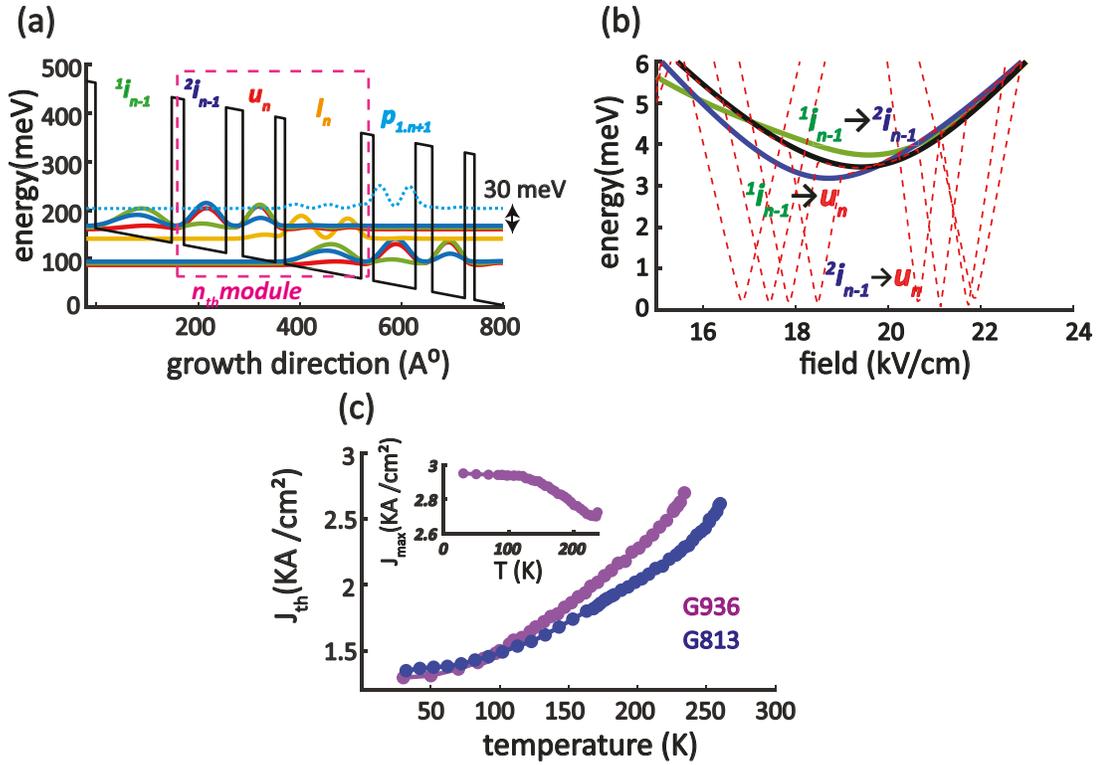

Fig.3. (a) probability density functions of subband states in a THz QCL with two injectors and direct phonon depopulation. (b) anticrossing graph for G936. The growth parameters starting from the injection barrier are **33**, 64.3, **19**, (59, 29, 59), **24**, 83.3. The bold fond indicates barriers, and the red ink indicates the doped region with $n_{3D}$=1.7x10$^{17}$/cm$^3$. The number of modules is chosen for ~10 $\mu$m of the active region. (c) comparing experimentally measured $J_{th}-T$ between G813 and G936 and $J_{max}-T$ for G936 as inset. Both designs have a lasing frequency of ~ 4 THz.

Our goal in this study is not to present the most optimized three-well scheme but rather a preliminary exploration, which can shed some light on the intermodule transport and provide a fair comparison with some designs based on 30% barriers. For instance, If the performance of G813 was limited by the tradeoff between transparency of the injection barrier and intermodule leakage, a three-well scheme with a similar injection barrier thickness leading to the same $J_{th}$ and $J_{max}$ should in principle, show a lower rate of increase in $J_{th}-T$ as compared to G813. Under this circumstance, the design direction would be to gradually reduce the thickness of the injection barriers to achieve a higher gain than G813. In this approach, a three-well design G936 was explored.

The anticrossing graph for G936 is shown in Fig. 3(b). The $J_{th}-T$ with $J_{max}-T$ as inset for G936 is plotted in Fig. 3(c). G936 achieved approximately the same $J_{th}$ as G813, and reached $T_{max}=235\,K$, the highest reported $T_{max}$ for a three-well scheme to this date (previously 200 K [32]). However, the $J_{th}-T$ shows a faster increase compared to G813. The drop of $J_{max}-T$ suggests that this design is lifetime limited, and the higher increasing rate of $J_{th}-T$ suggests thermal leakage. Further exploration is needed to investigate the impact of possible parasitic channels such as $|p_{n+1}\rangle$ that has a smaller energy separation (~30 meV) with injector doublet (Fig. 3(a)) than that in G813. More importantly, a systematic study is required to investigate if a reduction in *IFR* scattering in three-well schemes with reduced barrier composition (such as 25%) can offer advantages compared to the two-well schemes with barriers ≥ 30%.

**Conclusions**

In this paper, we achieved a new record in the maximum operating temperature of THz QCLs, $T_{max}=261K$. Considering the need for higher barrier compositions to suppress parasitic channels, optimizing the growth conditions to achieve a more defect-free interface might improve the temperature performance.

In addition, transport modeling improvement should narrow the gap between the simulated and the experimentally measured $T_{max}$. We have provided several designs in this paper, including a three-well scheme with $T_{max}=235K$ that could assist further theoretical and experimental investigations.

**Methods**

Methods are included in the supplemental file.

**Acknowledgments**

A.K would like to thank NextNano Inc. for providing a free evaluation license.

**Disclosures**

The authors declare no conflicts of interest.

**Contributions**

A. K. developed the simulation and optimization codes, designed and fabricated the lasers, and performed the measurements. MBE growth and related material characterization for samples whose IDs beginning with "G" were done by M.C. T under the supervision of Z. R. W . MBE growth and related material characterization for samples whose IDs starting with "VA" and "VB" were done by S.J.A and J. R A. K wrote the paper with editing help from all authors. Q. H. supervised the project.

**Funding**

This work is supported by the National Science Foundation (NSF) and the Natural Sciences and Engineering Research Council of Canada (NSERC). This work was performed, in part, at the Center for

**Supplemental information**

**Terahertz semiconductor laser source at -12 ⁰C**

Ali Khalatpour[1†*], Man Chun Tam[2*], Sadhvikas J. Addamane[3], John Reno[3], Zbig Wasilewski[2], Qing Hu[1†]

[1]*Department of Electrical Engineering and Research Laboratory of Electronics, Massachusetts Institute of Technology, Cambridge, MA 02139, USA;* [2]*Department of Electrical and Computer Engineering, University of Waterloo, 200 University Ave W, Waterloo, Ontario N2L 3G1, Canada.* [3]*Sandia National* Laboratories, *Center of Integrated Nanotechnologies, Albuquerque, NM 87185-130, USA*

*Equal contributions

[†] Corresponding authors: akhalatpour@g.harvard.edu, qhu@mit.edu

# 1. Introduction

To perform a non-equilibrium green function (NEGF) simulation, we used commercially available software from NextNano Inc [1]. For accurate NEGF simulations, the conduction band discontinuity (CBD), the material band structure (electrons and phonons), the distribution of ionized impurities, and a statistical model for interface roughness need to be specified. The material parameters, including bandgaps and their temperature behavior, non-parabolicity, and phonon energies, have been well studied and can be obtained with certain confidence from the literature [2]. On the other hand, interface roughness (*IFR*) parameters depend on the material growth condition (by Molecular Beam Epitaxy in this work). They are mainly used as a fitting parameters to explain the broadening of intersubband transitions [3,4]. Quantitatively, the CBD defines the effective height of the barriers. Therefore, it describes the characteristics of resonant and photon-assisted tunneling and the spatial overlap between electronic states that determine electron transport. Though CBD is essential for transport analysis, there has been no unanimous agreement on its value for $Al_xGa_{1-x}As$/GaAs heterostructures [2,5-6].

a. **Conduction band discontinuity(CBD)**

Fig. S1(a) shows the two-well scheme's probability distribution function for electronic states. Here, CBD is the depth of the quantum wells. CBD directly affects the spatial overlap between electronic states and directly impacts the form factors for various scattering mechanisms, including electron-electron, electron-phonon, and electron-impurity scattering [3,7]. Due to a short module length in a two-well scheme, these overlaps are more sensitive to the CBD than designs with longer module lengths involving more than two wells. In particular, the oscillator strength between $|u_n\rangle$ and $|l_n\rangle$, which determines both the optical gain and thermally activated *LO*-phonon emission rate, has a significant impact on the maximum operating temperature $T_{max}$. In addition, for any design, accurate estimation of $f_{ul}$ is needed to determine the optimum doping to achieve the highest $T_{max}$ [8]. The coupling strength at $|i_{n-1}\rangle \rightarrow |u_n\rangle$ alignment, quantified by the anticrossing $\Omega_{inj}$ (in the unit of meV) depends on the CBD and impacts both the maximum current density ($J_{max}$) and $T_{max}$. The lasing frequency can be determined through the energy of subbands and the broadening effects of scattering mechanisms and $\Omega_{inj}$ [7]. Therefore, a good value of CBD should provide a reasonable estimation of $J_{max}$, temperature performance, and lasing frequency.

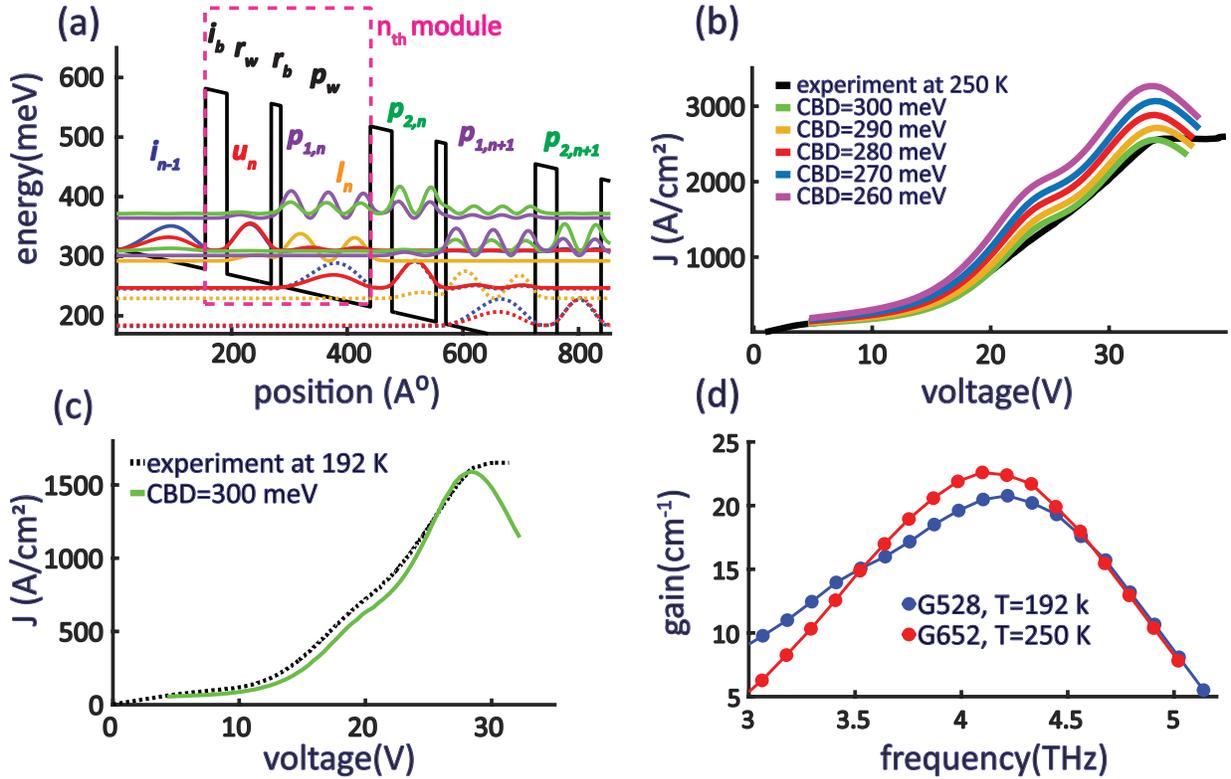

Fig. S1. (a) compares the simulated V-J with various CBD values to experimentally measured values. (b) G652 at $T_{max} = 250\ K$. (c) G528 at $T_{max} = 192\ K$. (d) compares the simulated gains between G652 and G528 at their corresponding $T_{max}$.

There are two significant limitations to using $J_{max}$ to estimate CBD. The maximum current density not only depends on the CBD but also depends on the doping density. The doping of a grown MBE wafer can only be determined with a ± 10% accuracy using measurement techniques such as C-V measurements. The limitation of using lasing frequency to estimate the CBD comes from its dependence on gain-broadening processes, $\Omega_{inj}$, and the impact of ionized donors on electronic states (Poisson effects), which are all functions of CBD. Using the mean-field approximation, the effect of ionized donors can be modeled as a position-dependent perturbation (Hartree potential) to the Hamiltonian of electronic states. The Poisson effect could be substantial for designs in which electrons are spatially remote from ionized donors [3]. In this case, the coupled Schrödinger-Poisson equation needs to be solved iteratively, but it also requires the band populations as an input. In a two-well scheme, the impact of ionized donors on electronic states due to Poisson effects is reduced as electronic states and ionized donors are spatially close. Since the Hartree potential acts as a perturbation to conduction band edges, this effect is relatively smaller for designs with tall barriers (~30% Al concentration) than low barriers (~15% Al concentration) for the same doping levels. The linewidth for THz QCLs is ~ 1 THz, leading to a strong dephasing rate of ~ 4 meV [7]. Therefore, one strategy to break the interdependency of lasing frequency on the CBD is to use designs with low injection coupling, $\Omega_{i,u} \approx 1-1.5 \text{meV}$. In this case, though the device will suffer from a low $T_{max}$ due to poor injection, the gain broadening due to $\Omega_{inj}$ is negligible. In the weak coupling regime, the impact of ionized donors on electronic states due to Poisson effects is reduced as the ground state, which primarily resides in the well where the dopants are, is the most populated subband, and thus the impact of ionized donors on the remote bands (here upper lasing levels) is minimal. In this limit, the lasing frequency can be used without the knowledge of transport to reasonably estimate CBD. In [9], this strategy was used to infer a value of ~ 300 meV for $Al_{0.3}Ga_{0.7}As/GaAs$, which corresponds to 72% conduction band offset.

To explore the validity of the suggested CBD for $Al_{0.3}Ga_{0.7}As/GaAs$, the best strategy is to compare designs that have widely different injection coupling (and, therefore $J_{max}$) and $T_{max}$. In this direction, we have chosen two designs with largely different $\Omega_{inj}$ which resulted in a significant difference in their corresponding $J_{max}$ and $T_{max}$ [9]. These two designs are grown in the same MBE chamber and within a short time from one another, which ensures similar doping calibration, background impurity, and interface roughness parameters. These designs are labeled G652 and G528 and are shown in Table S1.

Table S1: THz QCLs used for model training. Layer sequence starts from the injection barrier. Bold denotes $Al_{0.3}Ga_{0.7}As$ barriers separating GaAs quantum wells. The red color indicates the doped well, with a volume doping of $1.5\times10^{17}$ cm$^{-3}$ in the central 30-Å region. The number of modules is chosen for ~10 μm of active region, and the actual MBE growth thicknesses are listed.

| Wafer | Lasing frequency | $T_{max}$ | $\Omega_{inj}$ | Layer sequence (Å) |
|---|---|---|---|---|
| | THz | (K) | (meV) | |
| G528 | ~ 4THz | 192 | ~1.56 | **38.2**, 78, **17.1**, 157.8 |
| G652 | ~ 4 THz | 250 | ~2.67 | **33.9**, 72.2, **18.8**, 145 |

Fig. S1(b) shows the simulated voltage-current density (V- J) and its comparison with experimental data for G652 for various values of CBD. Here, the simulated voltage values are scaled by a constant so the simulated bias for $J_{max}$ overlaps with the experimental value. For the measured $J_{max}$, NEGF simulations suggest that the CBD=300 meV predicts the current density accurately at the nominal doping. Fig S1(c) shows that CBD=300 meV also predicts the $J_{max}$ for G528 at its nominal doping.

Since the CBD significantly impacts the $f_{ul}$, a relatively close estimated gain for two different designs at their corresponding $T_{max}$ indicates accuracy in the CBD estimation. Fig. S1(d) shows the simulated unsaturated gain for both wafers with CBD=300 meV. Based on this data, a similar gain level for G652 and G528 can be noticed at their $T_{max}$. Adding 20% uncertainty in estimating the doping densities can expand CBD to 275-310 meV range. In our MBE growth, the Si cell temperature is at about 1273K during QCL growth. Keeping the Si cell at a temperature within 1 deg gives us accuracy on Si flux of ~ 4%. However, in this work, Si cell was stabilized to 0.1 deg. Considering the lasing frequency of G528 and the higher bound on the temperature stability of our Si flux, our best judgment for CBD reduces to $292\pm8$ meV. As shown in Fig S2(a), the CBD= $292\pm8$ meV also predicts the $J_{max}$ of G813 with a $T_{max}$=260 K. The value of 289 meV for CBD has been recently used in studying two-photon absorption in GaAs/AlGaAs quantum well waveguides [6].

If we linearly increase the suggested range for x=0.3 to x=0.35, we expect that $335 \pm 8$ meV that also well predicts the $J_{max}$ for $Al_{0.35}Ga_{0.65}As/GaAs$ heterostructures.

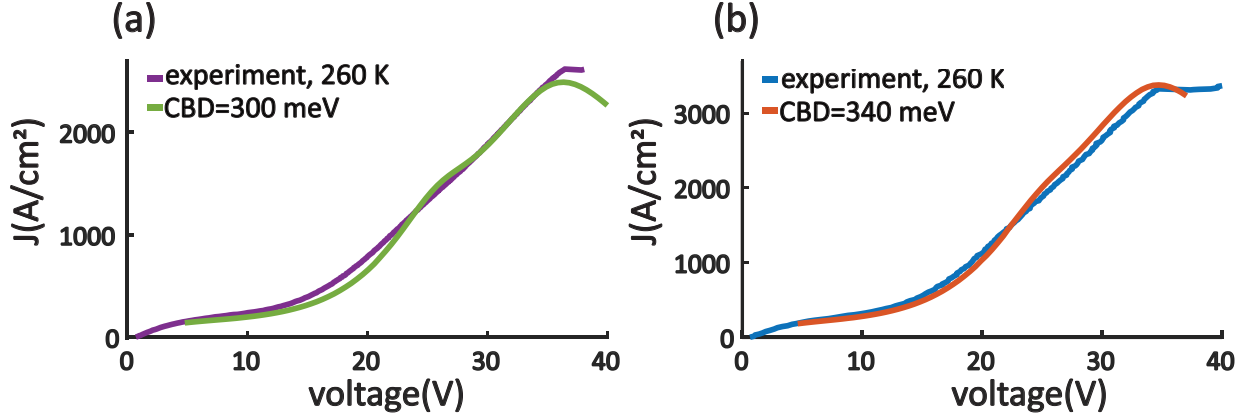

Fig. S2. a comparison between the simulated V-J compared to experimentally measured values. (a) G813 at $T_{max} = 260\ K$ with CBD=300 meV. (b) G902 (based on 35% Al barriers) at $T_{max} = 260\ K$ with CBD=340 meV.

Fig. S2(b), shows the simulation $J_{max}$ for G902, which is based on $Al_{0.35}Ga_{0.65}As/GaAs$ heterostructures and its comparison with experimentally measured values at $T_{max} = 260\ K$. Here, the suggested range $335 \pm 8$ meV can also predict the $J_{max}$.

b. **Interface Roughness Scattering**

One approach to model interface roughness (*IFR*) scattering in THz QCLs is to use the statistical description of a rough surface [3]. Randomly distributed uneven surfaces are characterized by the height distribution $h_\Delta$ and the height correlation length function $f_\Lambda$. The $h_\Delta$ determines the spread of heights away from an ideal interface plane and is often regarded as gaussian or exponential. The correlation function $f_\Lambda$ describes the extent to which knowledge of the height at one point on the surface does, on average, determines the height at some point $\Lambda$ away. In our simulation, we used an exponential model, $f_\Lambda = \Delta^2 e^{-\frac{|r|}{\Lambda}}$. Here. $\Lambda$ is the correlation length and $\Delta$ (step height) is the root mean square (RMS) of height over the surface. To assign the most predictive *IFR* parameters, we have used the $J_{th}$ of G652 at T=80 K as a reference.

Fig. S3 shows the experimental V-J of G652 along with the simulated values at 80 K. In the V-J plot, the onset of a shoulder-like feature is at $|i_{n-1}\rangle \rightarrow |l_n\rangle$ alignment, which is marked in the plot with a green dot. Since we have not included the free carrier loss in the metal contacts and the active region and the radiation loss from the laser facets (mirror loss), the transparency current ($J_{tp}$) is defined at gain G=0. Therefore, the threshold gain ($G_{th}$) can be obtained through $G_{th} = \frac{\alpha_m + \alpha_w}{\Gamma}$ in which $\alpha_m$ is the mirror loss, $\Gamma$ is the mode confinement factor in the cavity ($\sim 1$ here), and $\alpha_w$ is the waveguide loss due to metal contacts and free carrier loss. We also expect $\alpha_w$ to be a temperature-dependent parameter which is ~10 cm$^{-1}$ at low temperatures [10] and rises to ~ 20 cm$^{-1}$ at higher temperatures [11]. The inset in Fig. S3(b) shows the simulated gain for $\Delta = 0.8 \overset{0}{\text{A}}$ and $\Lambda = 80 \overset{0}{\text{A}}$ in which the simulated $J_{tp}$ and $J_{th}$ (assuming $\frac{\alpha_m + \alpha_w}{\Gamma} \sim 10\ cm^{-1}$) are marked with colored dots. The experimentally observed $J_{th}$ for G652 is marked with a horizontal dashed line for reference. We notice the NEGF simulation slightly overestimates the current in the plateau region as compared to the experimental values, the estimated $J_{th} \sim$ 1800 A/cm² is reasonably close to the observed value of $J_{th} \sim$1650 A/cm².

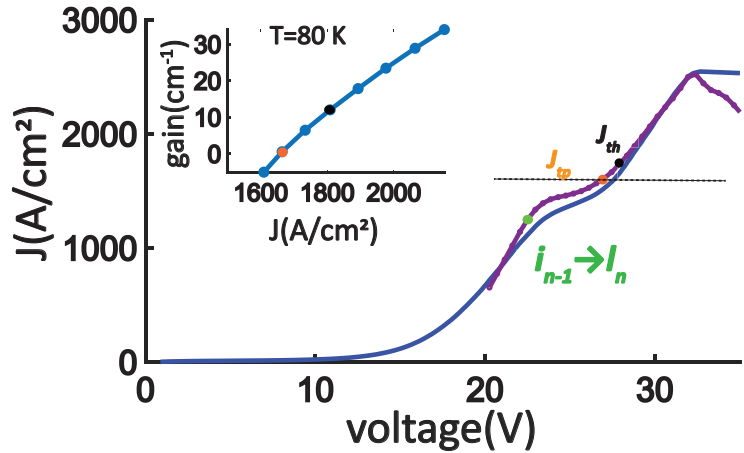

Fig. S3. voltage-current of G652 at 80 K. The experimental values are with a solid line, and the simulated values are with a dotted line. The inset shows the simulated gain for the proposed *IFR* parameters.

Though we have tried to extract a set of predictive *IFR* parameters, we emphasize that these parameters should be only used for structures that are grown in the same MBE chamber and with the same composition. Therefore, relative improvements of gain in various designs with the same compositions can be trusted. However, using other barrier compositions or designs with more barriers will doubt the predicted gain. As presented in the main file, *IFR* scattering has negligible impact on $J_{max}$. Therefore our analysis to determine CBD values based on $J_{max}$ remains valid.

### c. Distribution of ionized impurities

In [12], the segregation of Si dopants in Molecular Beam Epitaxy (MBE) grown GaAs quantum wells was investigated using current-voltage characteristics of quantum well infrared photodetectors. In the most predictive model, the dopant's position profile follows an asymmetric exponential spread with a long tail in the growth direction (here, the radiation barrier between $|u_n\rangle$ and $|l_n\rangle$). We believe such a profile is the most applicable model for ionized dopants in THz QCL. In this study, we used a similar profile for ionized doping profile $D_z$ when grown at 605 °C:

$$p_z^{z_i} = \begin{bmatrix} e^{\frac{z-z_i}{12 A^0}}, z \leq z_i \\ e^{-\frac{z-z_i}{3 A^0}}, z \geq z_i \end{bmatrix}, D_z = N \sum_{z_i} p_z^{z_i}$$

Here N is a normalization constant so $\int D_z dz = n_{2D}$ and $n_{2D}$ is the sheet doping density. The $p_z^{z_i}$ at $z_i = 0$ and $D_z$ are shown in Fig S4(a)-(b).

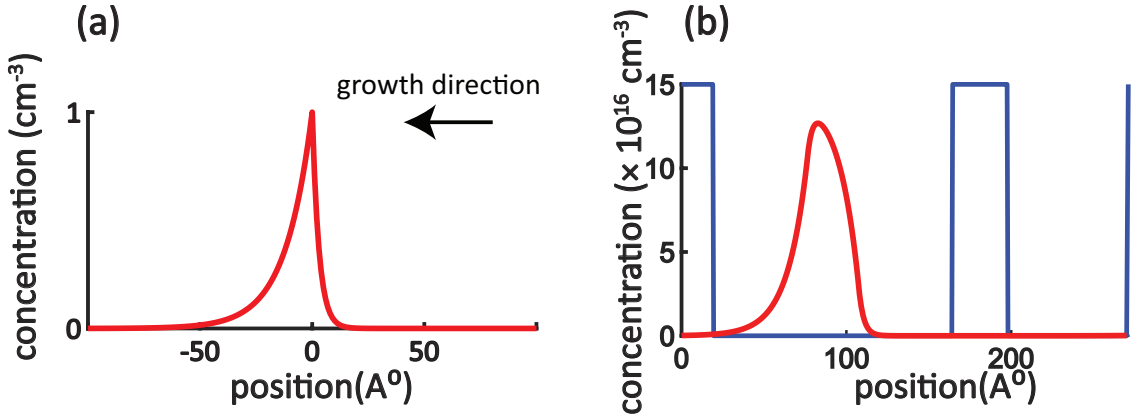

Fig. S4. The distribution of ionized dopants used in this work. (a) distribution of $p_z^0$. (b) $D_z$

### d. Non-parabolicity

Here we use a 3-band model that includes the impact of all the valence bands [3]. The energy-dependent effective mass of subband $i$ with energy $E_i$ is $m_e^*(E_i, z)$ and is defined as

$$\frac{m_0}{m_e^*(E_i,z)} = 1 - 2F + \frac{2}{3}\frac{E_p}{E_i - E_{lh}} + \frac{1}{3}\frac{E_p}{E_i - E_{so}}$$

Here we define $E_{so} = E_{lh} - \Delta_{so}$, $E_{lh}(z) = CBD - E_g - E_f z$ in which $E_f$ is the applied field and CBD is the conduction band discontinuity (CBD=0 is set for GaAs). Here

$$E_g^{Al_xGa_{1-x}As} = 1.517 + 1.403x - \frac{(1-x)\alpha_{GaAs}T^2}{\beta_{GaAs}+T} - \frac{x\alpha_{AlAs}T^2}{\beta_{AlAs}+T}$$

$E_p = 28.8(1-x) + 21.1x$ eV, $\Delta_{so} = 0.341(1-x) + 0.28x$ eV, $F = 1.94(1-x) + 0.48x$

For GaAs and AlAs we have used $\alpha$ =0.54 meV/K and $\beta$ =204 K, $\alpha$ =0.885 meV/K, $\beta$ =530 K, respectively. In this model, we set the CBD as a temperature invariant parameter, and the impact of temperature on band structure is modeled only through the change in effective masses. Considering the strong dependency of $J_{max}$ on CBD as shown in Fig. S1(b) and the fact that no significant change in the $J_{max}$ was observed in this experiment, using a temperature-independent CBD is justified.

## 2. Trade-offs involved in direct phonon depopulation scheme

Fig S5(a) shows the probability density functions of subband states in a THz QCL with the two wells scheme. The energy separation between $|i_{n-1}\rangle$ and $|p_{1,n}\rangle$ (denoted as $E_{i,pp}$) can be made high to reduce a possible leakage from $|i_{n-1}\rangle$ to $|p_{1,n+1}\rangle$ [9]. As the energy separation between localized states in a quantum well increases by decreasing the width of the well, $E_{i,pp}$ is mainly determined by the width of the phonon well (largest well) in which those states are localized. In this section, we show a subtle trade-off between reducing the intermoule-leakage (by increasing the $E_{i,pp}$

) and $f_{ul}\tau^u_{LO-em}$, where $\tau^u_{LO-em}$ is the lifetime due to nonradiative *LO*-phonon emission. Fig. S5(b) shows the figure of merit $f_{ul}\tau^u_{LO-em}$ for three out of four designs shown in Table S2 but with varying radiation well thickness ($r_W$) as labeled in Fig. S5(a). Here we also assume thermal distribution for subbands with a lattice temperature of 250 K and fixed electron temperature of 320 K for the upper lasing level. The details of this calculation can be found in [13]. For ease of comparison, $f_{ul}\tau^u_{LO-em}$ is normalized to that of G552. These designs have similar doping and barrier composition, ensuring a similar impact $\tau^u$ due to *imp* scattering and *IFR* scattering. In addition, these designs have similar $f_{ul}$, and $J_{max}$. As *e-e* scattering depends on both doping and $J_{max}$ [7], the impact of e-e scattering should be similar in those designs. The experimental results are summarized in Table S2.

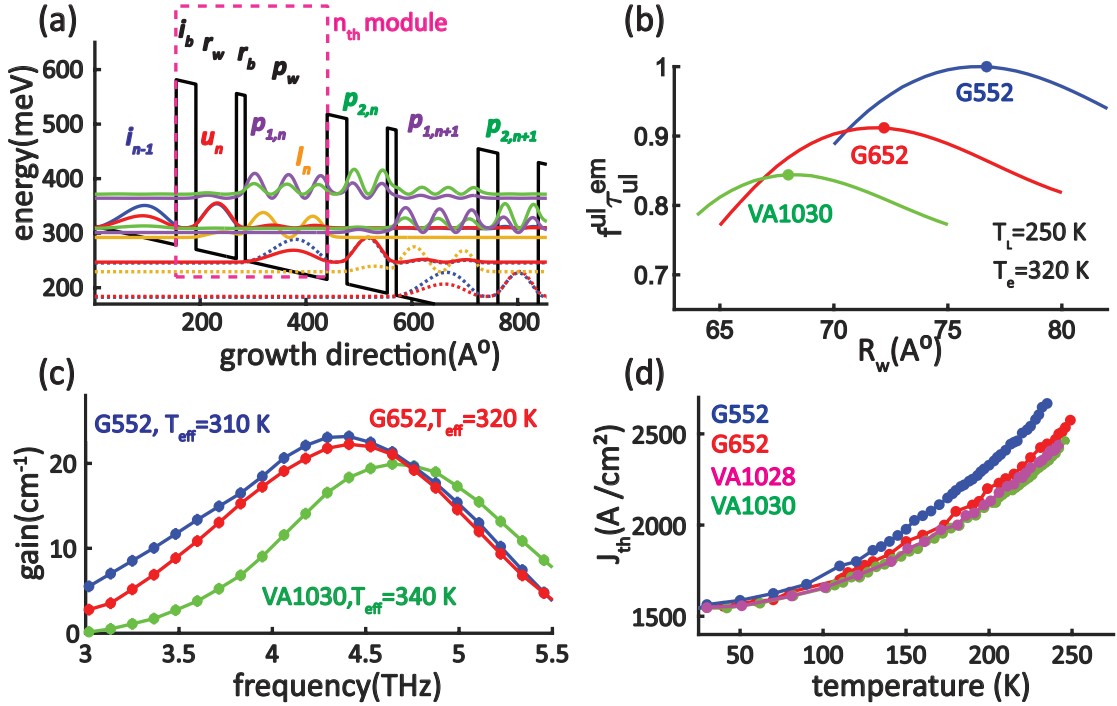

Fig.S5. (a) an example of a subband probability distribution function in a two-well scheme. Here The injection barrier, radiation well, radiation barrier, and phonon well are labeled $i_b, r_w, r_b, p_w$ respectively. (b) Calculated $f_{ul}\tau^u_{LO-em}$ for the wafers listed in Table S2 assuming an effective electron temperature Te=320 K for the upper lasing level and lattice temperature Tl =250 K. (c) unsaturated gain and effective electron temperature for upper lasing level calculated using NEGF. (d) experimentally measured $J_{th}-T$.

Table S2. The layer sequence starts from the injection barrier. Bold denotes Al$_{0.3}$Ga$_{0.7}$As barriers separating GaAs quantum wells. The underline indicates the doped well, with a volume doping of $1.5 \times 10^{17}$ cm$^{-3}$ in the central 30-Å region. The number of modules is chosen for ~10 μm of the active region. Here G552 and G652 were previously reported in [9].

| label | $T_{max}(K)$ | $f_{ul}$ | $E_{i,pp}(meV)$ | $J_{max}$ (A/cm2) | Layer sequence (Å) |
|---|---|---|---|---|---|
| G552 | ~235 | ~0.3 | ~53 | ~2660 | **33.2**, 76.7, **17.8**, <u>155</u> |
| G652 | ~250 | ~0.3 | ~57 | ~2600 | **33.9**, 72.2, **18.8**, <u>145</u> |
| VA1028 | ~242 | ~0.3 | ~58 | ~2450 | **33.7**, 71.5, **18.8**, <u>143</u> |
| VA1030 | ~245 | ~0.3 | ~61 | ~2470 | **33.3**, 68.0, **19.8**, <u>137</u> |

As shown in Fig S5 (b), designs with a narrower phonon well yielding an increaseed $E_{i,pp}$ will reduce the maximum $f_{ul}\tau^u_{LO-em}$. In addition, increasing $E_{i,pp}$ results in a higher effective electron temperature for $|u_n\rangle$ as the energy separation between $|l_n\rangle$ and $|g_n\rangle$ increases as compared to E$_{LO}$~36 meV in GaAs, resulting in excess kinetic energy after electrons relaxation from $|l_n\rangle$ to $|g_n\rangle$. This increase in effective electron temperature further reduces $f_{ul}\tau^u_{LO-em}$ by shortening the $\tau^u_{LO-em}$. This intuitive picture can be verified using non-equilibrium Green's function (NEGF) simulations. Fig S5 (c) shows the NEGF simulation at T= 250 K for the same designs. As expected, designs with higher $E_{i,pp}$ have higher effective subband temperature for $|u_n\rangle$ (T$_{eff}$) and reduced gain. However, contrary to the NGEF simulation's prediction, G552 had inferior temperature performance as compared to G652 though the latter benefited from a better Molecular Beam Epitaxy (MBE) growth quality [9]. In other words, if there are no other possible leakage channels, NEGF simulations favor designs with wide phonon wells in which the energy separation between $|l_n\rangle$ to $|g_n\rangle$ is closer to E$_{LO}$~36 meV in GaAs. A closer look at the measured $J_{th} - T$ presented in Fig S5 (d) shows a higher rate of increase in $J_{th}$ temperature for G552 when compared to G652, suggesting unaccounted leakage channels in the NEGF simulations for designs with a wide phonon well.

On the other hand, the $J_{th}-T$ of VA1030 shows no reduction of leakage compared to VA1028 (nominally the same as G652 but grown in a different MBE machine). Though it is hard to compare designs developed in separate MBE chambers, the experimental data suggest a fundamental trade-off involved in maximizing $f_{ul}\tau^u$ and suppressing inter-module leakage. In addition, a higher gain predicted by NEGF simulations for wider phonon wells cannot be experimentally observed. This factor was considered in our optimization scheme for the designs in Table 1 of the main file.

## 3. Gain degrading factors in G652

To further understand the impact of *imp*, *IFR*, and *e-e* scattering, we introduced a linear scale for the strength of each process in the NEGF simulations to evaluate the sensitivity of $J_{max}$ and optical gain (G) to those processes. In addition, since there are uncertainties in the model due to screening for the *e-e* and *imp* scattering, such scaling also determined the sensitivity of gain calculations to our assumptions. Fig. S6 (a)-(b) shows the simulated gain for various scaling of *imp* scattering ($s_{imp}$) and *e-e* scattering ($s_{ee}$), and Fig. S6(c) shows the calculated gain with and without *IFR* scattering. Though not shown here, the variation in simulated maximum current density $J_{max}$ is $\leq 3\%$ for all values of $s_{imp}$ and $s_{ee}$. The maximum current density through the injector barrier is qualitatively described by the Kazarinov-Suris expression [14,15].

$$J_{max} \sim \frac{eN}{2\tau_u} \frac{\left(\frac{\Omega_{i_{n-1},u_n}}{h}\right)^2 \tau_\| \tau_u}{1+\left(\frac{\Omega_{i_{n-1},u_n}}{h}\right)^2 \tau_\| \tau_u} \quad (1)$$

Here, $\tau_u$ is the upper state lifetime, $\tau_\parallel$ is the dephasing time between the injector and upper level, $\Omega_{i_{n-1},u_n}$ is the injection anticrossing in the unit of meV, h is the Planck constant, and N is the total electron sheet density of the two levels. Resonant tunneling limited transport corresponds to

$$J_{max} \sim \frac{eN}{2}\left(\frac{\Omega_{i_{n-1},u_n}}{h}\right)^2 \tau_\parallel, \text{ if } \left(\frac{\Omega_{i_{n-1},u_n}}{h}\right)^2 \tau_\parallel \tau_u \ll 1 \quad (2)$$

whereas lifetime-limited transport corresponds to the case

$$J_{max} \sim \frac{eN}{2\tau_u}, \text{ if } \left(\frac{\Omega_{i_{n-1},u_n}}{h}\right)^2 \tau_\parallel \tau_u \gg 1 \quad (3)$$

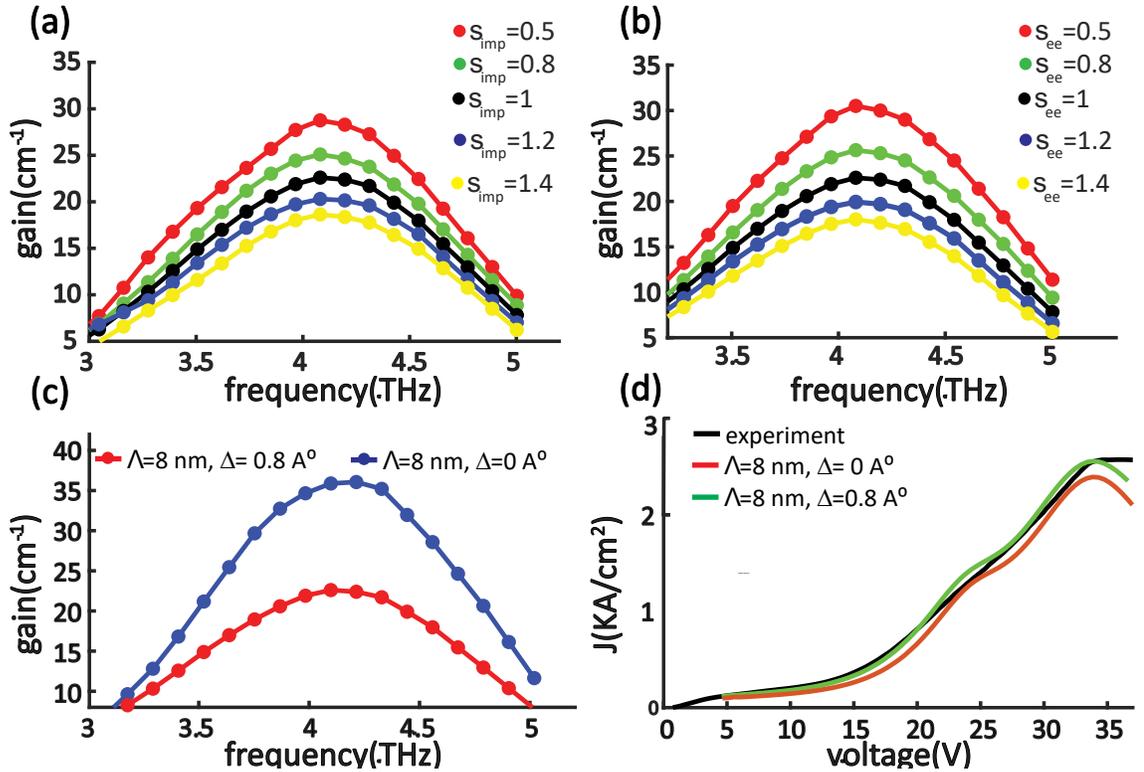

Fig.S6. estimated gain for various imp and e-e scattering scaling for G652 at 250 K. (a) *imp* scattering. (b) *e-e* scattering. (c) simulated unsaturated optical gain for G652 with and without *IFR* scattering at T=250 K. Here $\Delta$ and $\Lambda$ are the step height and correlation length and describe the statistical description of interface roughness [2]. (d) simulated voltage-current for G652 with and without *IFR* scattering at T=250 K.

The insensitivity of J$_{max}$ to all scattering mechanisms in simulation suggests that the $J_{max}$ in G652 is lifetime limited and the impacts of *IFR*, *imp*, and *e-e* scatterings are mainly gain broadening. However, as indicated in [16], the drop of $J_{max}$ with temperature in diagonal designs is a sign of switching between tunneling limited transport to lifetime limited transport, and it did not occur experimentally for G652. An explanation could be that the dephasings are slightly underestimated and $\left(\frac{\Omega_{i_{n-1},u_n}}{h}\right)^2 \tau_{\|}\tau_u \gg 1$ is not achieved in the experiment.

## 4. Gain Optimization procedure

It is instructive to consider the relative subband populations for G652 at 250 K to assess the importance of parasitic channels. This calculation is shown in Fig. S7(a). It appears the channel $|u_n\rangle, |i_{n-1}\rangle \to |p_{1,n+1}\rangle$ impacts the onset of population inversion (and therefore $J_{th}$) and the channel $|u_n\rangle, |i_{n-1}\rangle \to |p_{2,n+1}\rangle$ impacts the injection efficiency ($\eta_{inj} = \frac{2n_u}{n_i + n_u}$), both by depopulating the $|u_n\rangle$. Since there is a minimum number of barriers involved between $|u_n\rangle$ and $|p_{2,n+1}\rangle$, $\Omega_{u_n,p_{1,n+1}}$ can be considered as the highest bound on all parasitic channels. Therefore, by minimization of $\Omega_{u_n,p_{1,n+1}}$, all the parasitic channels can be minimized and can also lead to a reduction in $J_{th}$. The injection anticrossing $\Omega_{i_{n-1},u_n}$ mostly determines the transparency of the injection barrier and consequently $J_{max}$. If we choose $\zeta = \frac{\Omega_{i_{n-1},u_n}}{\Omega_{u_n,p_{1,n+1}}}$ as a figure of merit, we can improve $\eta_{inj}$ and $\tau^u$ without increasing the parasitic channel. Since both the injection barrier and the radiation barrier determine $\Omega_{u_n,p_{1,n+1}}$, while $\Omega_{i_{n-1},u_n}$ is mainly determined by the injection barrier, a higher $\zeta$ corresponds to a thicker radiation barrier (low $f_{ul}$) and thinner injection barrier (higher $\eta_{inj}$). Therefore, a lifetime limited transport can be observed by a gradual increase of $\zeta$ as compared to the one in G652. A suitable design strategy is then to use the upper bound of our estimation for CBD of 220 meV for GaAs/Al$_{0.25}$Ga$_{0.75}$As, 300 meV for GaAs/Al$_{0.3}$Ga$_{0.7}$As, and 340 meV for GaAs/Al$_{0.35}$Ga$_{0.65}$As heterostructures. With these values, $\zeta$ may be underestimated but it can be gradually increased while monitoring the rate of $J_{th}-T$ and $J_{max}-T$ as a figure of merit for suppression of parasitic channels and transparency of injection barrier. It is also worth mentioning that, in general, a higher $\zeta$ can be achieved with higher barrier composition, though it comes at the cost of higher *IFR* scattering.

Another critical design parameter that is highly sensitive to the CBD is the $|l_n\rangle \rightarrow |p_{2,n+1}\rangle$ alignment. Based on the NEGF simulations, co-aligning $|l_n\rangle \rightarrow |p_{2,n+1}\rangle$ and $|i_{n-1}\rangle \rightarrow |u_n\rangle$ can improve the optical gain as it introduces an extra channel of depopulation for $|l_n\rangle$. However, the uncertainty for CBD makes co-aligning $|l_n\rangle \rightarrow |p_{2,n+1}\rangle$ and $|i_{n-1}\rangle \rightarrow |u_n\rangle$ complicated from a design perspective. Multiple designs were explored in various barrier compositions to explore the suggested optimization scheme. The anticrossing graph for these designs and their comparison with G652 is shown in Fig. S7(b). As can be noticed, there is a gradual increase in $\Omega_{i_{n-1},u_n}$ while keeping $\Omega_{u_n,p_{1,n+1}}$ less than that of G652. The simulated gain and band populations using NEGF for these designs are shown Fig. S 7 (c)-(d). In G813, a double depopulation scheme is used by co-aligning $|l_n\rangle \rightarrow |p_{2,n+1}\rangle$ and $|i_{n-1}\rangle \rightarrow |u_n\rangle$.

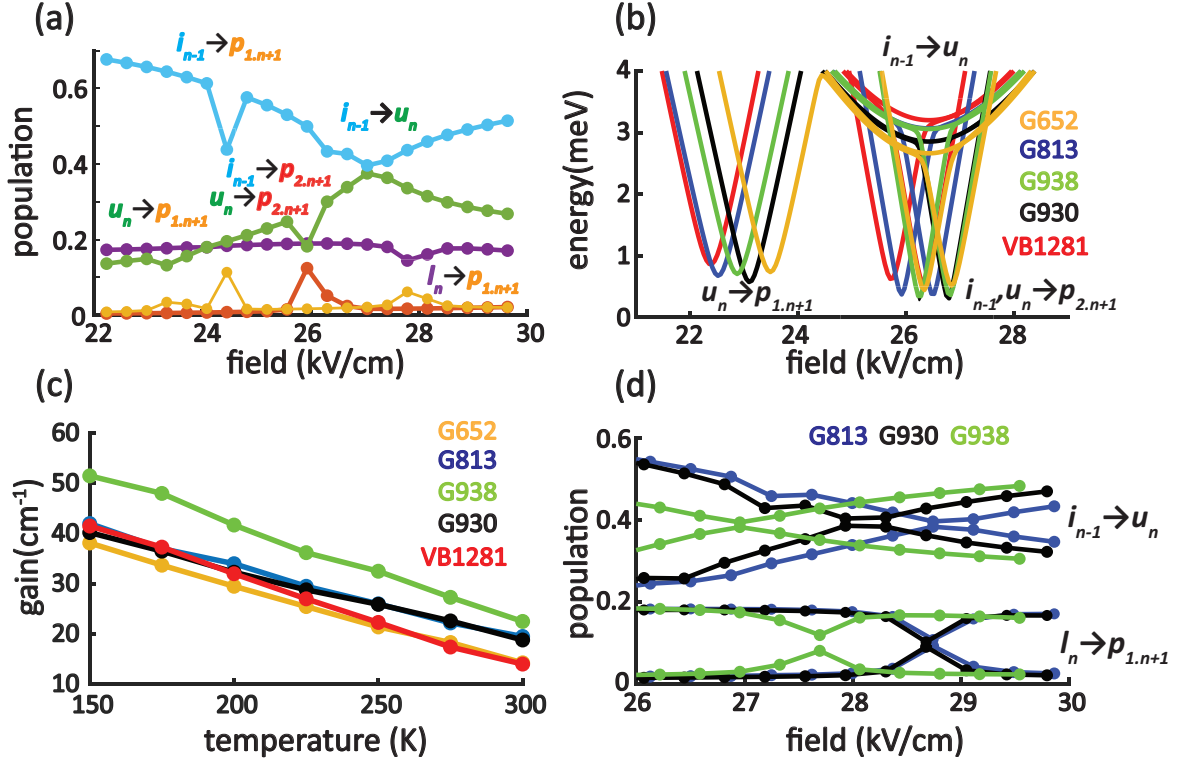

Fig. S7. (a) shows a simulated band population in G652. Here various subbands are indicated with the corresponding color. (b) anticrossing plot for different suggested designs. Here, for ease of comparison, the anticrossing graphs are shifted horizontally by a constant field to match the $|i_{n-1}\rangle \to |u_n\rangle$ alignment. Here the minimum of the parabola for $|i_{n-1}\rangle \to |u_n\rangle$ and $|u_n\rangle \to |p_{1,n+1}\rangle$ are $\Omega_{i_{n-1},u_n}$ and $\Omega_{u_n,p_{1,n+1}}$ respectively. A comparison between $\zeta = \dfrac{\Omega_{i_{n-1},u_n}}{\Omega_{u_n,p_{1,n+1}}}$ for various designs in reference to G652 can be visually inferred. (c) simulated gain vs temperature using NEGF. (d) simulated subband populations in proposed designs. Here $|i_{n-1}\rangle \to |u_n\rangle$ and $|l_n\rangle \to |p_{2,n+1}\rangle$ alignments are labeled for comparison. The definition of subbands is shown in part a of this figure.

## 5. Impact of doping profile

The position of dopants defines the overlap between the ionized impurities (here Si) and the probability density function of electronic states. This overlap determines the scattering rate with ionized impurities. The electronic states shown in Fig. 1(a) are plotted based on the extended scheme [7], in which no dephasing is introduced. In this scheme, $|u_n\rangle$ and $|i_{n-1}\rangle$ are delocalized equally and spread out in both phonon and radiation wells. However, including the dephasing processes increases the localization of the $|u_n\rangle$ in the radiation well and $|i_{n-1}\rangle$ in the phonon well. In addition, the overlap of $|u_n\rangle$ and ionized impurities depend on the oscillator strength. Intuitively, the probability density of $|l_n\rangle$ and $|i_n\rangle$ have maxima in the center of the phonon well. Therefore, some improvement may be possible by moving the ionized dopants away from the center and toward the radiation barrier. However, this comes at the cost of an increase in dephasing time for $|u_n\rangle$, but less severe for, diagonal designs due to a lower overlap of $|u_n\rangle$ with the ionized impurities. Therefore, determining the optimum doping position requires a complete transport analysis considering all dephasing processes. It should be noted that the doping scheme optimization assumes a smaller spread of dopants compared to the width of the phonon well. Though we used the dopant distribution shown in Fig S4 (b), it is instructive to compare our model against a uniform distribution. If the dopant distribution is uniform across the well, either by design or thermal diffusion, there should be no difference in the transport by various doping schemes. To test those two scenarios, two designs labeled as VB1278 and VA1186 were experimentally explored. Within the growth variation, VB1278 and VA1186 are identical designs but with two different doping profiles. The details of layers thicknesses and doping are shown in Table S3. Here, VA1186 is doped closer to the radiation barrier (here on called LD scheme), and VB1278 is doped at the center of phonon well doped (CD scheme).

Fig. S8(a) shows our models' schematic of dopant profiles. The behavior of $J_{th}$ with temperature is shown Fig. S8(b). As summarized in Table S3, The LD scheme had a marginally higher $T_{max}$ compared to CD scheme. In addition, the behavior of $J_{th}$ with temperature is noticeably changed. A higher rate of increase of $J_{th}$ with temperature was measured in CD scheme. This result suggests that the dopant positions can impact the transport and change the parasitic channel alignments, possibly due to Poisson effects. To explore this further, Fig. S8(c) shows the simulated gain and current density for VA1186 and VB1278. Here and along with the optimization scheme described in the main file, VA1186-B is also shown in the same plot for comparison purposes, but no experiments were pursued. Simulation suggests that the maximum gain and $J_{max}$ occur at the same bias in VA1186. On the other hand, the maximum $J_{max}$ and gain for VB1278 occur at different biases, suggesting some parasitic channels. To understand this further, the simulated band populations are shown in Fig. S8(d). This simulation suggests that the $|i_{n-1}\rangle \to |p_{2,n+1}\rangle$ occurs at a lower bias than $|i_{n-1}\rangle \to |u_n\rangle$ and could explain a higher rate of increase in $J_{th}$ with temperature. It is also worth noting that VA1186 relatively achieved ~ 10 K improvement compared to VA1028 (Table S2), suggesting the higher simulated gain was experimentally achieved. However, further, improvement is possible, as shown in Fig. S8(c) for VA1186-B.

Table S3. The layer sequence starts from the injection barrier. Bold denotes $Al_{0.3}Ga_{0.7}As$ barriers separating GaAs quantum wells. The phonon will is indicated by a parenthesis and the red ink indicates the doped region. The number of modules is chosen for ~10 μm of active region. The $n_{3D}$ (cm$^{-3}$) is $1.5 \times 10^{17}$.

| label | $T_{max}$ | Layer sequence (Å) |
|---|---|---|
| VB1278 | ~246 | **34.7**, 71.4, **21.4**, (55,37,56) |
| VA1186 | ~252 | **34.7**, 71.4,**22.4**, (28,36,84) |
| VA1186-B | NA | **32**, 71.4, **22.4**, (28,36,84) |

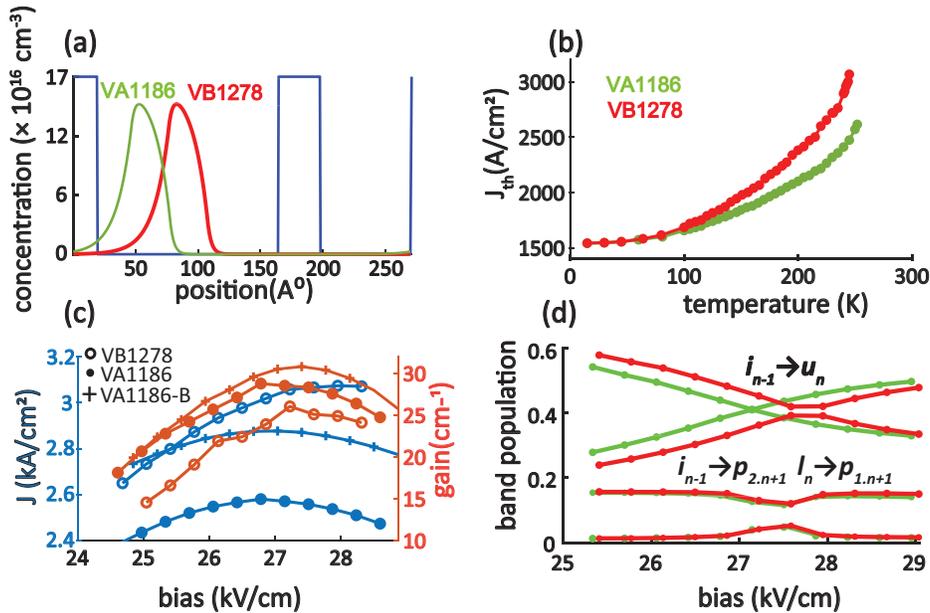

Fig. S8. (a) a model of dopant distribution for VA1186 and VB1278. (b) experimentally measured $J_{th} - T$. (c) simulation of gain and current density for VB1278 and VA1186, and VA1186-B . (d) simulation of band populations for VA1278 and VA1186.

Fig. S9 shows the simulation for LD, CD doped scheme for G938 at 260 K. The details of G938 with LD scheme were discussed in the main file. Here, the simulation predicts a similar level of relative gain improvement in the LD scheme compared t CD scheme.

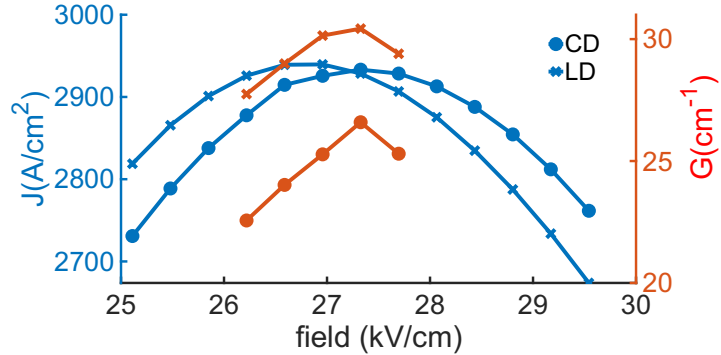

Fig. S9. simulation of gain and current density for LD, CD doping scheme for G938

## 6. Molecular Beam Epitaxy

All the quantum cascade laser (QCL) structures whose sample IDs begin with "G" were grown in a Veeco GEN10 molecular beam epitaxy (MBE) system at the University of Waterloo. The superlattices (SLs) were grown by using a Veeco SUMO v3 Ga cell, and a Veeco 60cc conical effusion Al cell for all structures. The Ga and Al cells were calibrated on a regular basis by performing high-resolution X-ray diffraction (HRXRD) analysis on dedicatedly grown AlAs-GaAs structures and extracting growth rates from those. To ensure precise control over the epilayers, in prior to each QCL growth, those group III fluxes were verified with an ion gauge, as well as by growing test layers of AlAs and GaAs on a separate GaAs (001) substrate and analyzing *in situ* optical reflectance oscillations [17]. Fig. S10(a) illustrates the $\omega$–$2\theta$ scan of G902 along with its corresponding dynamical simulation. The SL period variations are calculated from the observed broadening of the XRD satellite peak widths. We assume that the SL period varies slowly throughout the stack (e.g., due to Ga flux drift) and the full-width-half-maximum (FWHM) of those peaks approximately follows $\text{FWHM} = \sqrt{w_0^2 + (2.36\sigma\omega)^2}$, where $w_0$ is a constant related to instrumental broadening and the finite thickness of the stack, $\omega$ is the angular distance of the peak from the main 0[th] order reflection of SL, and $\sigma$ is the estimated standard deviation of period thickness. For Gaussian distribution, 2.36 is given by the ratio of the FWHM to $\sigma$. The FWHMs were extracted by fitting the peaks to a Gaussian function, then least-square-fit on the series of FWHMs was performed to extract $w_0$ (~7 arc sec) and $\sigma$ for each QCL. Fig. S10(b) shows the $\omega$–$2\theta$ scan of G652, G813, G902, G930 and G938 along with their $\sigma$ and $T_{max}$. The total growth duration of each THz QCL can last for approximately 20 hours, which makes effusion cell stability a critically essential factor. For instance, G902 has the best flux stability among all QCLs in this report, indicated by its narrowest FWHM found in its higher-order SL reflections from HRXRD measurements which gives its $\sigma$ close to ±0.07%. For G652, G813, G930, and G938, however, analysis of their HRXRD scans indicates that the cells were not as stable throughout the growth, with the most significant variation for the SL module thickness (G938) close to ±0.27%.

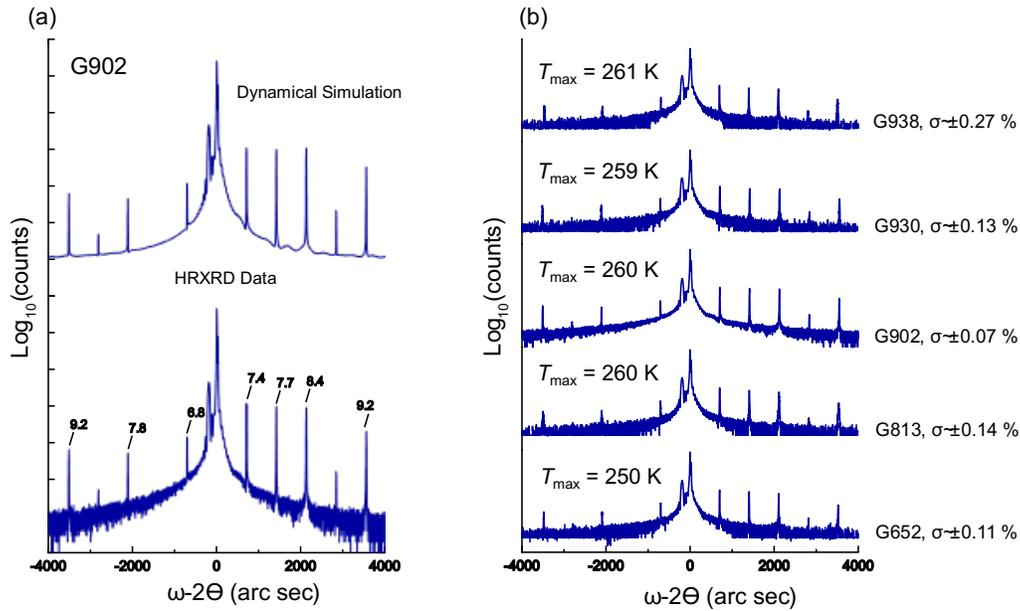

Fig. S10. (a) HRXRD (004) reflection measurement of G902. The dynamical simulation curve has been shifted up for clarity. Peak labels show the FWHM in arc seconds (note the log scale). The FWHM values are obtained by fitting the individual peaks with a Gaussian distribution. (b) HRXRD (004) scans for all five structures on a logarithmic scale. The scans are displaced by factors of 107 for clarity. The $\sigma$ values quoted are the estimated standard deviation in the superlattice period throughout the stack.

QCL sample IDs starting with "VA" or "VB" were grown at Sandia National Laboratories using MBE. VA samples were grown on an EPI1240 system, and VB samples were grown on a Veeco Gen20 system. Both reactors were equipped with Veeco Sumo© effusion cells for gallium, 60cc conical cells for aluminum and a 5cc dopant cell for silicon. The growth rates for Ga and Al were calibrated using reflection high-energy electron diffraction (RHEED) measurements and beam flux measurements (BFM) were done to verify good III-V ratios before each growth. The substrate temperature was measured using a pyrometer. The temperature of the Si cell was adjusted to obtain desired doping levels based on Hall measurements performed on calibration samples grown before device growth. Separate GaAs (AlAs)/AlGaAs superlattice structures were grown on both reactors and their periods were measured using XRD to verify growth rates. This also keeps the growth process consistent between the two reactors and ensures that VA and VB QCLs are comparable. The growths were carried out on 4" (100) GaAs substrates. First, the native oxide on the GaAs substrates was desorbed by thermal annealing at 630°C. The substrate temperature was then reduced to 600 °C and maintained there for the rest of the growth. Post-growth, each QCL was analyzed using XRD to verify period thickness and crystal quality. Fig. S11 shows $\omega - 2\theta$ scans for QCL structures described in this manuscript: VA1028, VA1030, VA1186, VB1278 and VB1286.

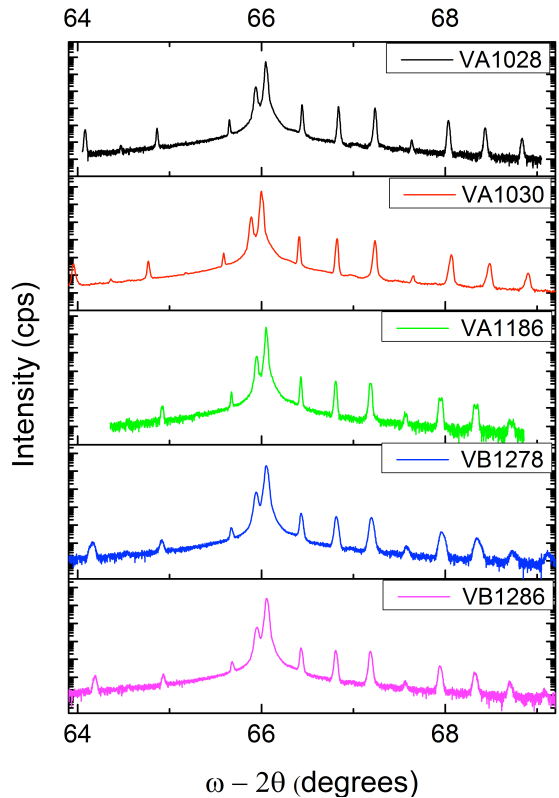

Fig. S11. HR-XRD (004) scans in logarithmic scale

## 7. Fabrication

From the MBE wafers, samples of typical size 1.5×1.5 cm$^{-2}$ were cleaved. In addition to the cleaved MBE sample, a slightly larger n+ GaAs substrate (from AXT, Inc.) is necessary; this substrate is named the "receptor," as the MBE growth is transferred to it during the fabrication process. Ta(10 nm)/Au(250 nm) was evaporated onto both the MBE sample and receptor substrate in a ~3×10$^{-7}$ Torr vacuum at a rate of 1 Å/s. The MBE sample and receptor were then bonded together using a metal-metal thermocompression at 300°C and 4 MPa of force for one hour in a vacuum. Then the samples were annealed at the same temperature but in atmospheric pressure nitrogen (N$_2$) and with no applied force for 45 minutes. Graphite spacers were used to distribute the force uniformly during bonding. Following wafer bonding, the native MBE substrate was lapped down to a thickness of ~100 μm. This remaining MBE substrate was then removed by wet etching in a 4:1 (by volume) solution of citric acid (1 g/mL) and 20% hydrogen peroxide (H$_2$O$_2$), which stops selectively at the Al$_{0.55}$Ga$_{0.45}$As etch-stop layer. The exposed etch-stop layer was selectively removed by concentrated (49%) hydrofluoric acid (HF). The top n+ contact was removed by a solution of H$_3$PO$_4$(1)/H$_2$O$_2$(1)/ H$_2$O(25). This completed the transfer of the gain medium epitaxial layer to the receptor substrate. The top contact/waveguide metal was then defined by a bilayer liftoff process (double coating of PGMI-SF5 and SPR700). The final top metallization layer was Ta(10 nm)/Au(300 nm), deposited under the same conditions as the first evaporation. The mesa was then defined through dry-etching in 1/7.5/28 sccm Cl2:SiCl4:Ar at 1 Pa, for which the top Au acts as a self-aligned mask. The dry etch recipe achieved vertical sidewalls thanks to forming a protective Si-based film during the etch. This layer had to be removed to facilitate successful device cleaving. The sidewall passivation was removed by immersion in buffered-oxide etchant (BOE). The receptor wafer was then lapped down to ~190 μm and metalized with Ti(10 nm)/Au(150 nm) at ~ 2×10⁻⁶ Torr vacuum with a 2 Å/s deposition rate. The overall fabrication process is shown in Fig. S12.

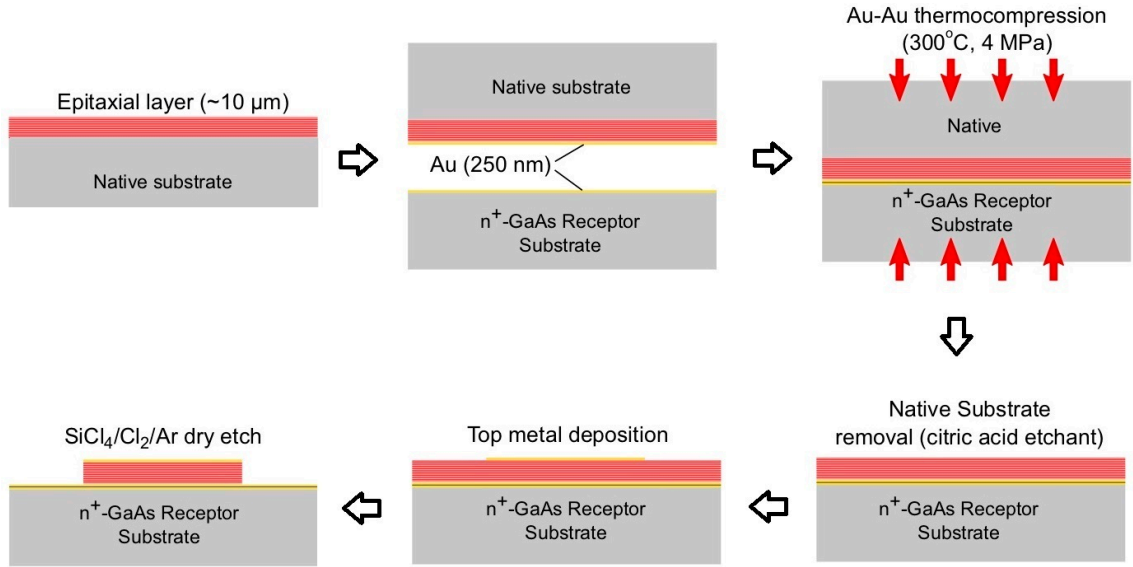

Fig. S12, Fabrication process of the THz QCLs in this paper.

## 8. Voltage-Current-Light measurement

The QCL devices were mounted in a closed cycle pulse-tubed helium cooler for measurement (model PT810, built by Cryomech Inc.). Characterization was performed using devices of approximate dimensions 1.3 mm × 150 $\mu$m. Lasers were biased using an Avtech model AVR-250B pulsed supply. Pulsing conditions were 400 ns at 500 Hz, corresponding to a 0.02% duty cycle. This extremely low-duty cycle is typically necessary to avoid device heating and obtain maximum lasing temperatures. Voltage and current were monitored using a digital oscilloscope (Pico 4824 from Pico Technology). Voltage measurements were made electrically parallel to the device under test. The current was indirectly sampled using an inductively coupled current sensor (current probe 711S from Integrated Sensor Technology). For optical detection, two instruments were used. For absolute power measurements, a calibrated power meter developed by Thomas Keating Ltd. was used. The average power measured by the meter was then divided by the duty cycle to obtain the peak pulsed power. For L-I measurements, the optical power output was measured using liquid helium (LHe) cooled Ga-doped Ge photodetector, built by Infrared Laboratories Inc. The uncalibrated power measurement was then scaled to match the absolute power measured by the Thomas Keating meter. Detector noise can mask a small optical signal, so it is essential to use a low-noise detector like the Ge:Ga photodetector for measurements near $T_{max}$. However, a downside of this detector is that it has a somewhat limited dynamic range. To overcome this, measurements at low temperatures were taken with a perforated anodized aluminum screen to prevent detector saturation. At high temperatures (near $T_{max}$), the screen was removed to obtain the highest possible sensitivity. Optical measurements were taken at intermediate temperatures with and without the screen. These intermediate measurements were used to scale the low-temperature measurements to the high-temperature measurements. Spectral measurements were performed using a Nicolet 8700 FTIR. Spectra were taken with 0.125 cm$^{-1}$ resolution.